\documentclass[useAMS,usenatbib]{mn2e}
\usepackage{array,multirow,pstricks,rotating,graphicx,amssymb,amsmath}
\usepackage{txfonts}
\usepackage{aas_macros}

\newcommand{\um}{\textup{ }\mu\textup{m}}
\newcommand{\qq}{\hspace{0.1cm}}
\newcommand{\cm}{\textup{ cm}}
\newcommand{\ff}{\textup{\tiny ff}}
\newcommand{\ev}{\textup{ eV}}

\newcommand{\amagat}{\textup{ amagat}}

\newcommand{\s}{\textup{ s}}
\newcommand{\K}{\textup{ K}}
\newcommand{\R}{\textup{ R}}
\newcommand{\eV}{\textup{ eV}}
\newcommand{\yr}{\textup{ yr}}
\newcommand{\Li}{\textup{Li}}
\newcommand{\LiH}{\textup{LiH}}
\newcommand{\HH}{\textup{H}}
\newcommand{\HD}{\textup{HD}}
\newcommand{\DD}{\textup{D}}
\newcommand{\He}{\textup{He}}
\newcommand{\e}{\textup{e}^-}
\newcommand{\kl}{\left[}
\newcommand{\kr}{\right]}
\newcommand{\lr}{\leftrightarrow}
\newcommand{\g}{\textup{ g}}
\newcommand{\const}{\textup{const}}
\newcommand{\conc}[1]{{\kl #1 \kr}}

\newrgbcolor{darkblue}{1.0 1.0 1.0}
\title{Rosseland and Planck mean opacities for primordial matter}
\author[M. Mayer \& W.J.\ Duschl]{Michael Mayer$^{1,3}$ and Wolfgang J.\ Duschl$^{1,2,4}$\\
$^1$ Institut f\"ur Theoretische Astrophysik, Tiergartenstr. 15,
69121 Heidelberg, Germany\\
$^2$ Steward Observatory, The University of Arizona, 933 N. Cherry
Ave., Tucson, AZ 85721, USA\\
$^3$ E-Mail: mm@ita.uni-heidelberg.de $^4$ Email:
wjd@ita.uni-heidelberg.de}

\begin{document}
\date{\today}
\pagerange{\pageref{firstpage}--\pageref{lastpage}} \pubyear{2004}

\maketitle

\label{firstpage}

\begin{abstract}
We present newly calculated low-temperature opacities for gas with a
primordial chemical composition. In contrast to earlier calculations
which took a pure metal-free Hydrogen/Helium mixture, we take into
account the small fractions of Deuterium and Lithium as resulting
from Standard Big Bang Nucleosynthesis. Our opacity tables cover the
density range $-16 < \log \rho \qq [\textup{g} \cm^{-3}] < -2 $ and
temperature range of $1.8 < \log T \qq [\textup{K}] < 4.6$, while
previous tables were usually restricted to $T > 10^3\,\mathrm{K}$.
We find that, while the presence of Deuterium does not significantly
alter the opacity values, the presence of Lithium gives rise to
major modifications of the opacities, at some points increasing it
by approximately 2 orders of magnitude relative to pure
Hydrogen/Helium opacities.
\end{abstract}

\defcitealias{1976ApJS...31..271C}{CT76}
\defcitealias{1986ApJ...302..590S}{SPS86}
\defcitealias{1991ApJS...76..759L}{LCS91}
\defcitealias{2004ApJ...600.1025H}{HLMT04}

\section{Introduction}
According to the Standard Big Bang Nucleosynthesis (SBBN), the most
abundant elements in the aftermath of the Big Bang were Hydrogen and
Helium with very small traces of Deuterium and Lithium. These
elements make up the so called POP {\sc iii} or
\textit{primordial\/} chemical composition. All the heavier elements
present today had to be produced in the course of the evolution of
several generations of stars.

In order to understand the physics of the early Universe, there is a
need of having appropriate and accurate material functions at hand,
in particular opacities for a SBBN chemical compositions.

There is a large body of literature available with opacity
calculations for metal-free (Z=0) H/He mixtures, starting with the
Paczynski I-VI mixtures \citep[][hereafter
\citetalias{1976ApJS...31..271C}]{1976ApJS...31..271C}, the
calculations of \citet[][hereafter
\citetalias{1986ApJ...302..590S}]{1986ApJ...302..590S} and
\citet[][hereafter
\citetalias{1991ApJS...76..759L}]{1991ApJS...76..759L}, and more
recently \citet[][hereafter
\citetalias{2004ApJ...600.1025H}]{2004ApJ...600.1025H}.

These calculations were restricted to temperatures between $10^3\K$
and $10^4\K$. Specific metal-free opacity sets have been calculated
by the OP project \citep[][$10^{3.75}\dots
10^{8}\K$]{1994MNRAS.266..805S} and OPAL \citep[][$10^3\dots
10^{8.7}\K$]{1994ApJ...437..879A, 1996ApJ...464..943I}. More
recently, OPAL calculations have been extended
\citep{2004MNRAS.348..201E} to $10^3\dots 10^{10} \K$ and
densities\footnote{$R=\rho/T_6^3$ is a parameter which replaces the
density in order to keep opacity tables in rectangular format when
spanning many decades in temperature (for more details see Appendix
\ref{Sect:tblfmt}); $T_6 = T / 10^6\K$} $-8 < \log \R < 7$ with the
main application for CNO enhanced opacities in stellar evolution.

All these opacity calculations, however, only considered pure
Hydrogen/Helium mixtures with no metals (mass fraction $Z=0$), but
different ratios $X/Y$ of the mass fractions of Hydrogen ($X$) and
Helium ($Y$). It has been argued that due to the assumed very small
abundances of Deuterium and Lithium these element do not play any
significant role in the opacities and thus in the evolution of POP
{\sc iii} objects.

Moreover, while there are good low-temperature opacities available
for a solar chemical composition
\citep[e.g.][]{2003A+A...410..611S}, zero-metallicity opacity
tabulations have been restricted to temperatures above $1000 \K$ so
far.

In this paper we present calculations of both, Rosseland and Planck
mean opacities for primordial matter including all the three
elements (Hydrogen, Helium, and Lithium) including Deuterium isotope
and present their absorption properties.

In Sect.\ \ref{Sect:matter} we discuss the quantitative composition
of primordial matter at a given temperature and density. We then
describe the relevant absorption processes (Sect.\
\ref{Sect:absproc}) and present our Rosseland and Planck mean
opacities (Sect.\ \ref{Sect:rossplanck}). Here, we also assess the
influence of Lithium on the opacity for different Lithium contents.
Before we compare our results to previous tabulations (Sect.\
\ref{Sect:compare}), we give analytic calculations for the molecule
formation timescale in Sect. \ref{Sect:chemequilibrium} in order to
estimate the time needed to reach chemical equilibrium which is one
of the underlying assumptions in our calculations. Finally, we
summarize our conclusions in Sect.\ \ref{Sect:conclusion}.

\section{Primordial Matter} \label{Sect:matter}

Primordial Matter as created in the SBBN consists of Hydrogen (H),
Helium (He), Deuterium (D), and Lithium (Li). Before
WMAP\citep{2003ApJS..148..175S}, the observed abundances of these
elements have been used to constrain the baryon-to-photon ratio. In
the framework of a $\Lambda$CDM cosmology WMAP provided the value
this parameter with high accuracy. While the abundances of H and He
were fairly well known for some time, the uncertainties in the
determination of the abundances of D and Li were significantly
reduced. Now, the D abundance as derived from cosmological
parameters is consistent with direct observations, while the
observed Li abundance still lacks a factor of 3 compared to the SBBN
results. This may indicate either systematic effects in the
observations or new physics \citep[for an in-depth discussion of
this issue, see][]{2004ApJ...600..544C}. We summarize the resultant
abundances according to SBBN and WMAP in Table \ref{tbl:abundances}
\citep[after][]{2004ApJ...600..544C} and take them as our {\it
fiducial\/} POP {\sc iii} mixture. For later comparison purposes we
define also a Li-free mixture in Table \ref{tbl:abundances}.

\begin{table}
\begin{center}
{\large
\renewcommand{\arraystretch}{2.0}
\begin{tabular}{l|c|c}
\hline \hline
 & {\large POP {\sc III}} & {\large Li-free}\\ 
\hline \hline
Y & $(0.2479\pm 0.0004)$ &  $(0.2479\pm 0.0004)$ \\
\hline
D/H & $(2.60^{+0.19}_{-0.17})\cdot 10^{-5}$ & $(2.60^{+0.19}_{-0.17})\cdot 10^{-5}$ \\
\hline
$^3\He/\HH$ & $(1.04^{+0.04}_{-0.04})\cdot 10^{-5}$ & $(1.04^{+0.04}_{-0.04})\cdot 10^{-5}$ \\
\hline
$^7\Li/\HH$ & \bf{$(4.15^{+0.49}_{-0.45})\cdot 10^{-10}$} & \bf{$0$} \\
\hline
\end{tabular}}
\end{center}
\caption{SBBN concordance abundances \citep{2004ApJ...600..544C}
from first-year WMAP results}\label{tbl:abundances}
\end{table}

\renewcommand{\multirowsetup}{\centering}
\begin{figure*}
\begin{center}
\includegraphics[width=\textwidth]{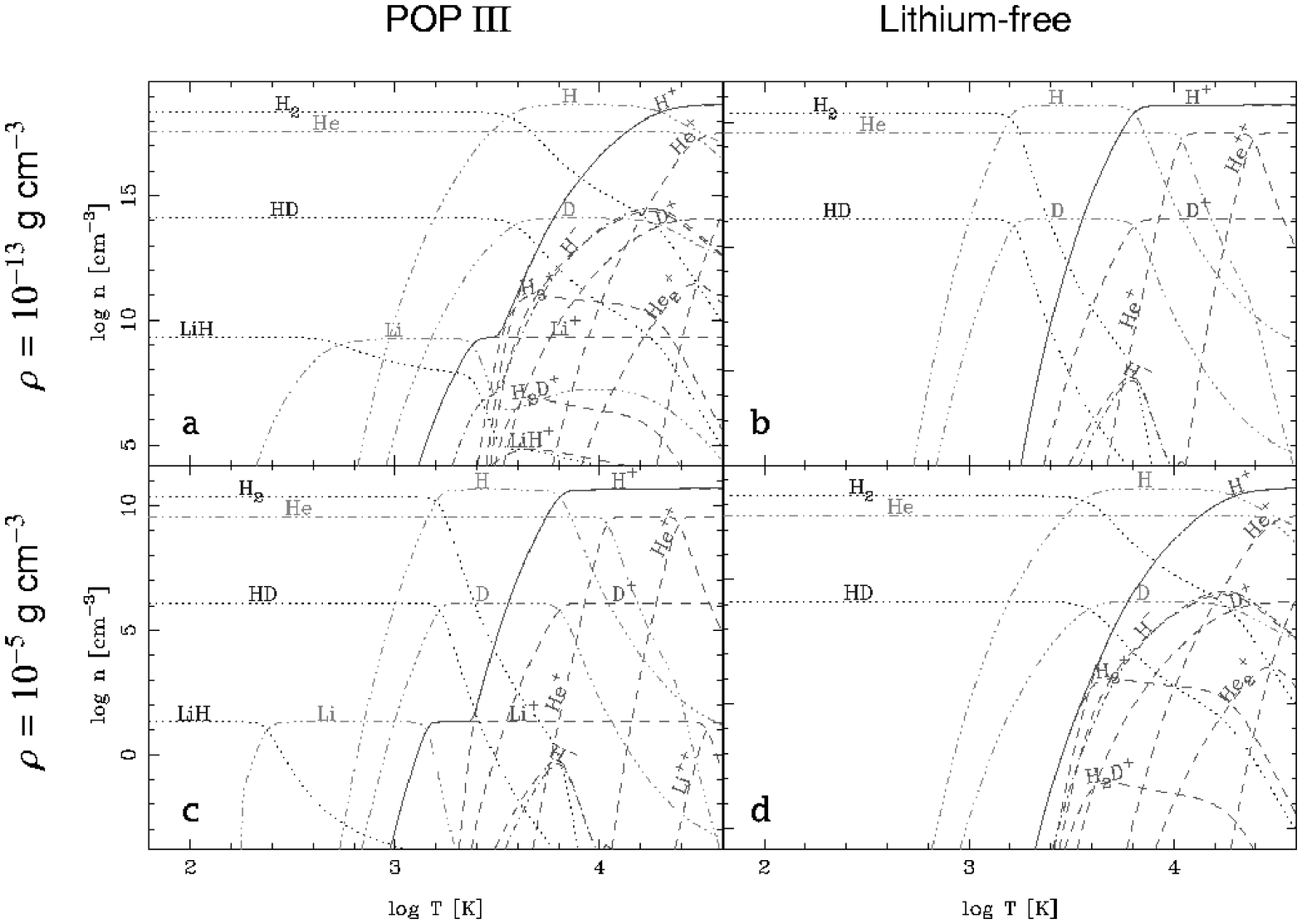}
\end{center}
\caption{The influence of the presence or absence of Li on the EOS
for two densities: {\bf (a)} $\rho=10^{-13}\g\cm^{-3}$, POP {\sc
iii}; {\bf (b)} $\rho=10^{-13}\g\cm^{-3}$, Li-free; {\bf (c)}
$\rho=10^{-5}\g\cm^{-3}$, POP {\sc iii}; {\bf (d)}
$\rho=10^{-5}\g\cm^{-3}$, Li-free. The different lines have the
following meanings: solid red line -- electron abundance; dotted
blue lines -- abundance of neutral molecules; dashed purble lines --
positive ions; dash-dotted black lines -- negative ions;
dash-triple-dotted green lines -- abundance of neutral atoms.}
\label{fig:liabundance}
\end{figure*}

\subsection{Equation of state (EOS)} \label{Sect:EOS}

We consider 20 species of primordial matter: electrons $\e$,
Hydrogen species $\HH^-$, $\HH$, $\HH^+$, $\HH_2$, $\HH_2^+$, and
$\HH_3^+$, Helium species $\He$, $\He^+$, $\He^{++}$, and $\He_2^+$,
Deuterium species $\DD$, $\DD^+$, $\HD$, and $\HH_2\DD^+$, and
Lithium species $\Li$, $\Li^+$, $\Li^{++}$, $\LiH$ and $\LiH^+$.

The computations assume chemical and thermodynamical equilibrium.
Equilibrium constants (see Table \ref{tbl:eqconst}) are used
whenever they are available in the literature, otherwise appropriate
Saha equations are taken.

For the partition functions the most recent $\HH_3^+$ partition
function $Q_{\HH_3^+}$ of \citet{1995ApJ...454L.169N}, for $\HH_2^+$
\citet{1994JQSRT...51..655S} is used. For $\HD$ and $\HH_2$ a
partition function for the equilibrium constant is calculated using
the energy levels for $\HD$ \citep{1997ApJ...490..76S}
and\footnote{Data from {\tt http://www.physast.uga.edu/ugamop/}}
$\HH_2$ . All other partition functions needed for the Saha
equations are approximated by the ground state statistical weight.

Combining all the equilibrium constants, there are 5 equations
utilizing mass conservation of all 4 elements and charge neutrality.
These equations are solved using an iterative procedure (see
Appendix \ref{App:EOS}).

\begin{table*}
\begin{center}
\renewcommand{\arraystretch}{1.5}
\begin{tabular}{l|l|l}
\hline
\hline
$\HH_2 \lr \HH+\HH $ & $K_{\HH_2}=\frac{\conc{\HH }\conc{\HH }}{\conc{\HH_2 }}$  & Saha $(U=4.478 \eV)$\\
$\HH^- \lr \HH+\e$ & $K_{\HH^-}=\frac{\conc{\HH} [ \e ]}{\conc{\HH^- }}$  & Saha $(U=0.7556 \eV)$\\
$\HH \lr \HH^++\e$ & $K_{\HH}=\frac{\conc{\HH^+} \conc{\e }}{\conc{\HH }}$ & Saha $(U=13.6 \eV)$\\
\multirow{2}*{$\HH_2^+ \lr \HH+\HH^+$} & \multirow{2}*{$K_{\HH_2^+}=\frac{\conc{\HH } \conc{\HH^+}}{\conc{\HH_2^+ }}$} & \citet{1994JQSRT...51..655S}\\
& &  $D_0 = 2.6507 \eV$ \citep{huberherzberg}\\
$\HH_2^+ + \HH_2 \lr \HH_3^+ + \HH $ & $K_{\HH_3^+}=\frac{\conc{\HH_3^+} \conc{\HH }}{\conc{\HH_2^+} \conc{\HH_2 }}$ & see App. \ref{Sect:h3plus} \citep[$D_0 = 0.7382 \eV$, ][]{1992A+A...255..453S} \\
\hline
\hline
$\HD \lr \HH+\DD $ & $K_{\HD}=\frac{\conc{\HH } \conc{\DD}}{\conc{\HD}}$ & Saha \citep[$U=4.5167 \eV$,][]{1997ApJ...490..76S} \\
$\DD \lr \DD^++\e $ & $K_{\DD}=\frac{\conc{\DD^+ } \conc{\e}}{\conc{\DD}}$ & $K_{\DD}=K_{\HH}$  (ass.) \\
$\HH_3^+ + \DD \lr  \HH_2\DD^+ + \HH$ & $K_{\HH_3^+,\DD}=\frac{\conc{\HH_2\DD^+} \conc{\HH }}{\conc{\HH_3^+ } \conc{\DD }}$ & \citet{1992A+A...255..453S},  $D_0 = 0.0439 \eV$ \\
$\HH_3^+ + \HD \lr \HH_2\DD^+ +\HH_2$ & $K_{\HH_3^+,\HD}=\frac{\conc{\HH_2\DD^+ } \conc{\HH_2}}{\conc{\HH_3^+ } \conc{\HD }}$ & \citet{1992A+A...255..453S},  $D_0 = 0.0120 \eV$\\
\hline
\hline
$\He \lr \He^++\e$ & $K_{\He}=\frac{\conc{\He^+ } \conc{\e }}{\conc{\He}}$  & Saha $(U=24.587 \eV)$\\
$\He^+ \lr \He^{++}+\e$ & $K_{\He^+}=\frac{\conc{\He^{++}} \conc{\e }}{\conc{\He^+ }}$  & Saha $(U=54.418 \eV)$\\
\multirow{2}*{$\He_2^+ \lr \He+\He^+$} & \multirow{2}*{$K_{\He_2^+}=\frac{\conc{\He } \conc{\He^+}}{\conc{He_2^+ }}$}  & \citet{1994JQSRT...51..655S}\\
& & $D_0 = 2.365 \eV$ \citep{huberherzberg}\\
\hline
\hline
$\Li \lr \Li^+ +\e$ & $K_{\Li}=\frac{\conc{\Li^+ } \conc{\e }}{\conc{\Li}}$ & $K_{\Li}=K_{\LiH^+,1}\cdot \left. K_{\LiH,2}\right/K_{\LiH,1}$ \\
$\Li^+ \lr \Li^{++}+\e$ & $K_{\Li^+}=\frac{\conc{\Li^{++} } \conc{\e }}{\conc{\Li^+}}$ &  Saha $(U=75.640 \eV)$ \\
$\LiH \lr \Li+\HH$ & $K_{\LiH,1}=\frac{\conc{\Li } \conc{\HH }}{\conc{\LiH }}$ & \citet{1996JQSRT...54..849S}\\
$\LiH \lr \LiH^+ + \e$ & $K_{\LiH,2}=\frac{\conc{\LiH^+ } \conc{\e}}{\conc{\LiH }}$ & \citet{1996JQSRT...54..849S}\\
$\LiH^+ \lr \Li^+ + \HH$ & $K_{\LiH^+,1}=\frac{\conc{\Li^+ } \conc{\HH}}{\conc{\LiH^+ }}$ & \citet{1996JQSRT...54..849S}\\
$\LiH^+ \lr \Li + \HH^+$ & $K_{\LiH^+,2}=\frac{\conc{\Li } \conc{\HH^+}}{\conc{\LiH^+ }}$ & \citet{1996JQSRT...54..849S}\\
\hline
\hline
\end{tabular}
\end{center}
\caption{Equilibrium constants}\label{tbl:eqconst}

\end{table*}

\subsection{Equilibrium $\HH_2^++\HH_2\leftrightarrow \HH_3^++\HH$} \label{Sect:h3pluseos}

Since the early work of \citet{1961JChemPhys...49..962P} it has been
clear that $\HH_3^+$ is important for temperatures of $2000 -
5000\,\mathrm{K}$
\citep{1984MolPhys...51..887T,1991JQSRT...45..57C,1992A+A...255..453S,1995ApJ...454L.169N}
as in this range it is the most abundant positive molecular ion (cf.
\citetalias{1991ApJS...76..759L}). The improvement of the
understanding of $\HH_3^+$ enhanced the reliability and enlarged the
high-temperature partition function which leads to an increase in
abundance.

Although the maximum abundance never reaches that of $\HH^-$ and $\HH_2^+$, it
influences the equilibrium abundances due to the charge neutrality
condition. A fit to the latest equilibrium constant using the latest
partition functions can be found in Appendix \ref{Sect:h3plus}.

\subsection{The influence of D and Li on the EOS}

Deuterium does not affect the equilibrium abundances of the Hydrogen
and Helium species given its small abundance and isotopic relation
to Hydrogen.

Lithium belongs to the group of Alkali metals. It therefore can
easily be ionized at comparably low temperatures. Fig.
\ref{fig:liabundance} shows Lithium providing free electrons at
comparably low temperatures (T$\approx 800\dots 3000$ K). These
enhance the $\HH^-$ but diminish the $\HH_3^+$ abundance in the
high-density limit, while there is no change in the $\HH^-$ and
$\HH_3^+$ abundances for lower densities, where only the electron
abundance is increased. Although being the least abundant species in
the mixture, it influences the equilibrium abundances of the
Hydrogen species.

Depending on the densities, both elements can form hydrides at low
temperatures ($T \lessapprox 1000\K$ and $T \lessapprox 300\K$,
respectively)

\section{Absorption processes} \label{Sect:absproc}

\begin{table*}
\begin{center}
\renewcommand{\arraystretch}{1.2}
\begin{tabular}{l|l|l|l}
\hline
\hline
{\large Thomson} & Th($\e$) &$\e + h\nu \rightarrow \e + h\nu'$ & \cite{2000asqu.book.....C} \\
\hline
\hline
\multirow{4}*{\large Rayleigh} & Ray($\HH_2$) & $\HH_2 + h\nu \rightarrow \HH_2 + h\nu'$ & \cite{1962ApJ...136..690D}\\\cline{4-4}
& Ray($\HH$) & $\HH + h\nu \rightarrow \HH + h\nu'$& \multirow{3}*{\citet{2000RaPC...59..185K}}\\
& Ray($\He$) & $\He + h\nu \rightarrow \He + h\nu'$& \\
& Ray($\Li$) & $\Li + h\nu \rightarrow \Li + h\nu'$& \\
\hline
\hline
\multirow{12}*{\large free-free} & ff($\HH^-$) & $\HH+\e +h\nu \rightarrow \HH + \e$& \citet{1987JPhysB...20..801B} using Fit of  \citet{1988A+A...193..189J} \\\cline{4-4}
& \multirow{2}*{ff($\HH$)} & \multirow{2}*{$\HH^+ + \e + h\nu \rightarrow \HH^+ + \e $} & \citet{1979rpa..book.....R} \\
& & &  $\quad$ using Gaunt-factors from \citet{1998MNRAS.300..321S}\\\cline{4-4}
& ff($\HH_2^+$) & $\HH^+ + \HH + h\nu \rightarrow \HH^+ + \HH$ & \citet{1994ApJ...430..360S}\\
& ff($\HH_2^-$) & $ \HH_2 + \e +h\nu \rightarrow \HH_2 + \e$ & \citet{1980JPhysB...13..1859B}\\
& ff($\HH_2$) & $\HH_2^+ + \e + h\nu \rightarrow \HH_2^+ + \e $ & $\sigma_{\ff}(\HH_2)=\sigma_{\ff}(\HH)$ (ass.)\\
& ff($\HH_3$) & $\HH_3^+ + \e + h\nu \rightarrow \HH_3^+ + \e $& $\sigma_{\ff}(\HH_3)=\sigma_{\ff}(\HH)$ (ass.)\\
& ff($\He_2$) & $\He_2^+ + \e + h\nu \rightarrow \He_2^+ + \e$& $\sigma_{\ff}(\He_2)=\sigma_{\ff}(\HH)$ (ass.)\\
& ff($\He_2^+$) & $\He^+ + \He + h\nu \rightarrow \He^+ + \He$ & \citet{1994ApJ...430..360S}\\
& ff($\He$) & $\He + \e + h\nu \rightarrow \He+ \e$ & \citet{1994MNRAS.269..871J}\\
& ff($\He$) &$\He^+ + \e + h\nu \rightarrow \He^+ + \e$& $\sigma_{\ff}(\He)=\sigma_{\ff}(\HH)$ (ass.)\\
& ff($\He^+$) &$\He^{++} + \e + h\nu \rightarrow \He^{++} + \e$ & $\sigma_{\ff}(\He^+)=\sigma_{\ff}(\HH)$ (ass.)\\
\hline
\hline
\multirow{9}*{\large bound-free} & bf($\HH^-$) & $\HH^- + h\nu \rightarrow \HH + \e$ & $\lambda < 1.6419\um$, \citet{1979MNRAS.187P..59W} using Fit of  \citet{1988A+A...193..189J}\\\cline{4-4}
&  \multirow{2}*{bf($\HH$)} & \multirow{2}*{$\HH + h\nu \rightarrow \HH^+ + \e$} & Method of \citet{1992QB809.G67......}\\
& & & $\quad$ Gaunt-Factor from \citet{1967ApJ...149..169M},\citet{1961ApJS....6..167K}\\\cline{4-4}
 & bf($\HH_2$) & $\HH_2 + h\nu \rightarrow \HH_{2,diss}^*$ & $h\nu > 15.4\ev$, \citet{1998ApJ...496.1044Y,2001ApJ...559.1194Y}\\
 & bf($\HH_2^+$) & $\HH_2^+ + h\nu \rightarrow \HH^+ + \HH $& \citet{1994ApJ...430..360S}\\
& bf($\He$) & $\He + h\nu \rightarrow \He^+ + \e$ & \citet{1967ZA.....66..185H}\\
& bf($\He^+$) & $\He^+ + h\nu \rightarrow \He^{++} + \e$ & \citet{1967ZA.....66..185H}\\
& bf($\He_2^+$) & $\He_2^+ + h\nu \rightarrow \He^+ + \He $& \citet{1994ApJ...430..360S}\\
& bf($\Li$) & $\Li + h\nu \rightarrow \Li^+ + \e$ & 2s and 2p state: \citet{1988JPhB...21.3669P}, Hydrogenic otherwise\\
\hline
\hline
\multirow{14}*{\large bound-bound} & bb($\HH_3^+$) & $\HH_3^+ + h\nu \rightarrow \HH_3^{+,*}$ & \citet{1996ApJ...464..516N}\\\cline{4-4}
& \multirow{3}*{bb($\HH_2$)} & \multirow{3}*{$\HH_2 + h\nu \rightarrow \HH_2^*$} & $A_{21}$ from \citet{1998ApJS...115..293W}\\
& & & energy levels from the Molecular Opacity Database\\
& & & $\quad$ UGAMOP ({\tt http://www.physast.uga.edu/ugamop/})\\ \cline{4-4}
& \multirow{2}*{bb($\HD$)} & \multirow{2}*{$\HD + h\nu \rightarrow \HD^*$} & $A_{21}$ from \citet{1982A+AS...50..505A}\\
& & &  energy levels from \citet{1997ApJ...490..76S}\\\cline{4-4}
& \multirow{3}*{bb($\LiH$)} & \multirow{3}*{$\LiH + h\nu \rightarrow \LiH^*$} & $A_{21}$ from \citet{1980JChPh..73.5584Z}\\
& & & with additions according to \citet{1997MNRAS.288..638B}\\
& & & $\quad$  energy levels from \citet{1996ApJ...458..397D}\\\cline{4-4}
& \multirow{2}*{bb($\HH$)} & \multirow{2}*{$\HH + h\nu \rightarrow \HH^*$} &\citet{1966atp..book.....W}\\
& & & Stark broadening from \citet{1999A+AS..140...93S}\\\cline{4-4}
& bb($\He$) & $\He + h\nu \rightarrow \He^*$ &NIST \citep{NIST}\\
& bb($\Li$) & $\Li + h\nu \rightarrow \Li^*$ &NIST \citep{NIST}\\
& bb($\Li^+$) & $\Li^+ + h\nu \rightarrow \Li^{+,*}$ &NIST \citep{NIST}\\
\hline
\hline
\multirow{14}{2cm}{\large Collision\\induced\\absorption} & \multirow{4}*{CIA($H_2/\HH_2$)} & \multirow{4}*{$\HH_2+\HH_2+h\nu \rightarrow \HH_2+\HH_2$} & $60\K<T<1000 \K$,  $10\cm^{-1}<\nu<14 000\cm^{-1}$ \\
& & & $\quad$ \cite{2002A+A...390..779B}\\
& & &  $1000\K<T<7000 \K$,  $20\cm^{-1} <\nu <20 000\cm^{-1}$ \\
& & & $\quad$ \cite{2001JQSRT..68..235B}\\\cline{4-4}
 & \multirow{6}*{CIA($\HH_2/\He$)} & \multirow{6}*{$\HH_2+\He+h\nu \rightarrow \HH_2+\He$} & $40\K<T<1000\K$, $40\cm^{-1}<\nu<15000\cm^{-1}$ \\
& & & $\quad$ RV $0\rightarrow 1$:\citet{1989ApJ...336..495B}\\
& & & $\quad$ RV Overtones: \citet{1989ApJ...341..549B}\\
 & & & $\quad$ RT : \citet{1988ApJ...326..509B}\\
& & & $1000\K<T<7000\K$, $25\cm^{-1}<\nu<200088\cm^{-1}$ \\
& & & $\quad$ \cite{2000A+A...361..283J}\\\cline{4-4}
& \multirow{2}*{CIA($\HH/\He$)} & \multirow{2}*{$\HH+\He+h\nu \rightarrow \HH+\He$} & $1500\K <T<10000 \K$, $50\cm^{-1}<\nu<11000\cm^{-1}$\\
& & & $\quad$\cite{2001ApJ...546.1168G}\\\cline{4-4}
& \multirow{2}*{CIA($\HH_2/\HH$)} & \multirow{2}*{$\HH_2+\HH+h\nu \rightarrow \HH_2+\HH$} & $1000\K<T<2500 \K$, $100\cm^{-1}<\nu<10000\cm^{-1}$\\
& & & $\quad$\cite{2003A+A...400.1161G}\\
\hline
\hline
\end{tabular}
\end{center}

\caption{Scattering and absorption processes} \label{tbl:absproc}

\end{table*}

In our calculations we take into account

\begin{itemize}
\item Thomson scattering;
\item Rayleigh scattering;
\item Free-free absorption;
\item Bound-free absorption including\\[-18pt]
\begin{itemize}
\item photoionisation and
\item photodissociation;\\[-18pt]
\end{itemize}
\item Bound-bound absorption; and
\item Collision-induced absorption.
\end{itemize}

\noindent For a detailed list of the processes considered including
references, see Table \ref{tbl:absproc}.

\subsection{Collision-induced absorption (CIA)}

CIA is basically the van der Waals interaction between a pair of not
neccessarily polar molecules or atoms. The interaction induces a
dipole momentum which leads to absorption. Although called
collision-induced absorption, this should not be confused with the
dipole moments created by an external field. CIA induced dipole
momenta are in this sense ''permanent''.  The usual timescale of
this interaction is of the order of nano-seconds, therefore we
expect broad and diffuse absorption spectra.

A general theory of CIA has already been developed in the 1950s
\citep[mainly
by][]{1957Physica...23..825K,1958Physica...24..347K,1959CJPhys...37..1187K}.
The first application to astrophysics has been done by
\citet{1969ApJ...156..989L} in the case of late-type stars. During
the last decade, this theory has been improved by semi-analytical
quantum mechanical computations obtained from first principles using
newly available dipole moments and an improved line-shape theory
\cite[e.g.][]{1989ApJ...336..495B}

The importance of CIA has been shown in the model atmospheres of
cool, low-metallicity stars \citep{1997A&A...324..185B} and cool
white dwarfs \citep[][]{2000A+A...361..283J}. It is also important
in the atmospheres of planets. Observational confirmation has been
put forward by \cite{2002ApJ...580.1070B} in the case of two
ultra-cool white dwarfs. It also has been shown to be important in
the zero-metallicity opacity calculation by
\citet{1991ApJS...76..759L} and \cite{2004ApJ...600.1025H}, the
first one still partially relying on the Linsky spectra.

In the cool, low-metallicity gas the importance of CIA arises
because there is no dust absorption. The main absorber is $\HH_2$
absorbing itself only due to quadrupole transitions which usually
are very weak compared to dipole transitions. Therefore CIA
introduces the possibility of more powerful absorption in these
environments. Due to the strong dependence of the absorption
coefficient on the number density of the contributing species
($\propto n_a n_B$, see Sect. \ref{Sect:database}) we expect this
absorption to be important in high density regions.

\subsection{Bound-bound absorption}

Bound-bound absorptions resulting from 8 species have been included
in our opacity calculations (see Table \ref{tbl:absproc}). While the
Einstein coefficients and energy levels are available in the
literature, the exact modeling of the lineshape itself proved to be
difficult. We tried a simple Doppler profile and a Voigt profile
\citep[following][]{1993JQSRT..50..635S} including Doppler
broadening and radiative damping for the Lorenz wings. In both cases
the contribution of the molecular lines to the Planck opacities
execeeds that of the continuum by several orders of magnitudes while
it does not considerably influence the Rosseland mean.

However, for a more realistic modeling of the lineshapes Stark
broadening for atomic lines (already considered for atomic H) and
collisional broadening for the molecular lines (already considered
for $\HH_2$ through CIA) have to be taken into account. We consider
Stark broadening for the first 3 series of atomic Hydrogen
\citep{1999A+AS..140...93S}. This is an important opacity source
even for the Rosseland mean as it is able to fill the ''valleys''
between the absorption edges of bound-free H absorption in the
monochromatic absorption coefficients.

All other bound-bound absorptions (except $\HH_3^+$, see below) are
treated according to the integrated line absorption coefficient
(App. \ref{Sect:integabs}) and are only considered for the Planck
mean. We verified this approximation by running a extensive
computation including Voigt profiles for the lines in both Rosseland
and Planck mean opacity calculation, but never reached differences
larger than 10\%.

Given the importance of line contributions, the Planck and Rosseland
averaging only yields approximative estimates of the true
monochromatic absorption. Even the line positions are subject to
change with respect to gas motions, etc. Hence we give two sets of
Planck opacities (see Section \ref{Sect:rossplanck}).

\subsection{$\HH_3^+$ bound-bound absorption}

$\HH_3^+$ influences the abundances of the other species in a
temperature range of $2000-5000\K$ (see Sect. \ref{Sect:h3pluseos})
as then it is the most abundant positive molecular ion.

The bound-bound $\HH_3^+$ line absorption has been shown by
\citetalias{2004ApJ...600.1025H} to have an effect of
low-metallicity, very low-mass stellar evolution. We use the line
list of more than 3 million lines \citep{1996ApJ...464..516N}, but
bin them to frequency intervals of $10 \cm^{-1}$. Our tests with
different bin sizes ranging from $200\dots 0.1 \cm^{-1}$ did not
produce significantly different results for the Rosseland mean so we
used this smoothed coefficients which we tabulated in the
$\left(\nu,T\right)$-plane. Without loosing accuracy, we save a
considerable amount of CPU time. Due to the huge number of lines,
$\HH_3^+$ line absorption appears as continuuous absorption. For
line absorption in the Planck mean, we use the integrated absorption
coefficient (see App. \ref{Sect:integabs}).

\subsection{Database} \label{Sect:database}

For all tabulated absorption coefficients bicubic splines are used
unless otherwise stated in Table \ref{tbl:absproc}. The tabulated
free-free absorption were interpolated by bicubic splines, too, but
extrapolated to low frequencies
\citep[cf.][$\kappa_{\nu,{\textup{\tiny ff}}}\propto
\nu^{-3}\left(1-\exp\left(-h\nu/k_\mathrm{B}T\right)\right)\propto\nu^{-2}$]{1994MNRAS.269..871J}.

The CIA data is taken from tables available on the Web\footnote{see
{\tt http://www.astro.ku.dk/}\~{\tt aborysow/programs/}} except for
H$_2$/He, where programs available from the same source are used to
calculate the roto-vibrational and roto-translational spectra. These
routines are used for temperatures lower than $1000\K$, while for
the high-temperature region ($1000\dots 7000\K$) the latest ab
initio data were applied.
\begin{figure*}
  \centering
  \includegraphics[width=\textwidth]{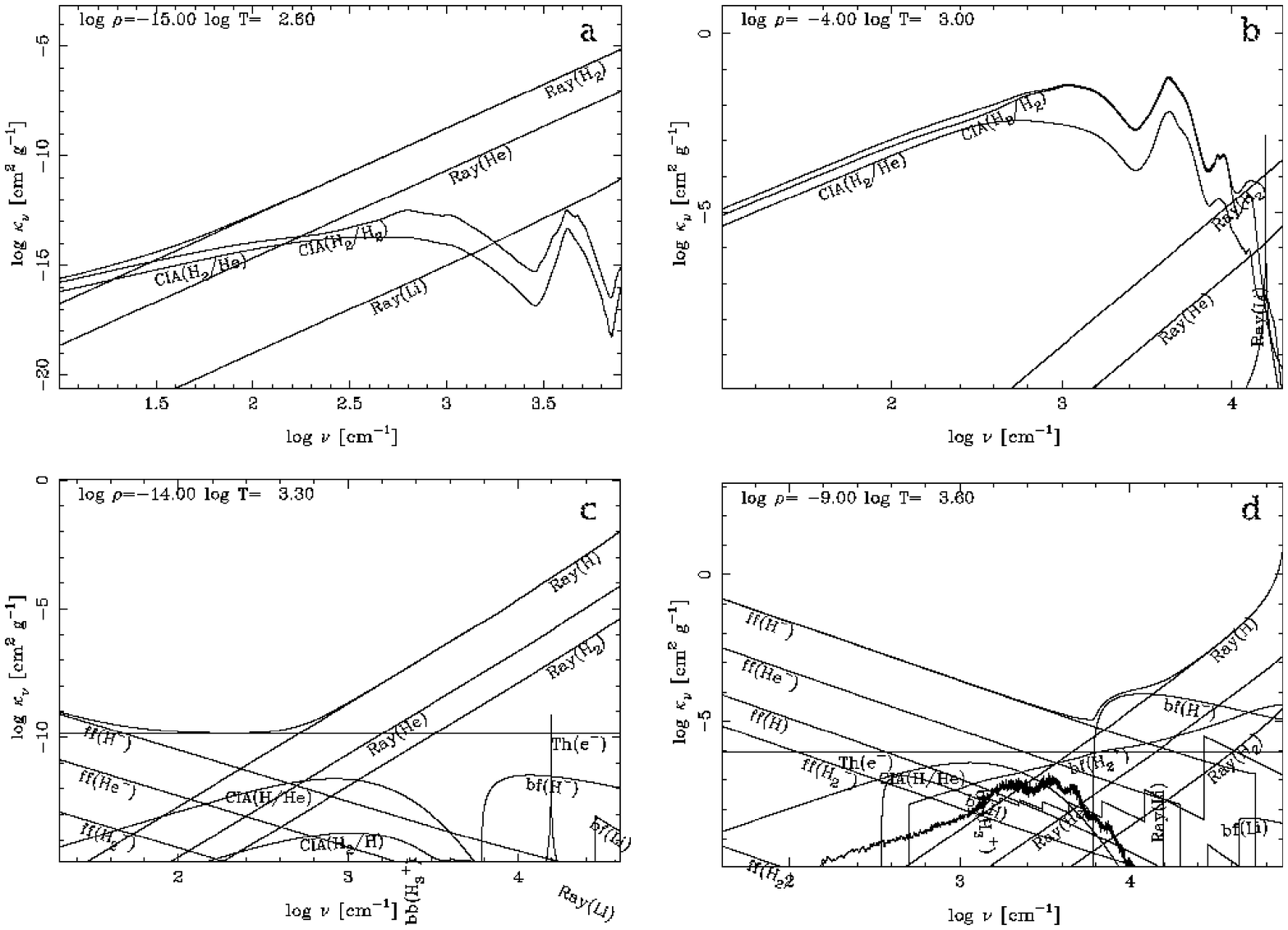}
  \caption{Absorption coefficient $\kappa_s(\nu)+\kappa_a(\nu)$ for different densities and temperatures: {\bf (a)} $\rho=10^{-15}\g\cm^{-3}$, $T=10^{2.6}\K$; {\bf (b)} $\rho=10^{-4}\g\cm^{-3}$, $T=10^{3}\K$; {\bf (c)} $\rho=10^{-14}\g\cm^{-3}$, $T=10^{3.2}\K$; {\bf (d)} $\rho=10^{-9}\g\cm^{-3}$, $T=10^{3.6}\K$. Also shown are the contributions of the different absorption processes.}
  \label{fig:abscoef1}
\end{figure*}
Extrapolation to non-tabulated low frequencies was done according
$\kappa_{\nu,\textup{\tiny CIA}}\propto \nu^2$. This extrapolation
is justified as for given temperature at low frequencies, the
absorption coefficient for CIA(A/B)
\citep[e.g.][]{1989PhRvA..39.2434M}
\[
\alpha(\omega)=\frac{4\pi^3}{3hc}n_An_B\omega g(\omega,T)\left(1-e^{-\frac{h\omega}{2\pi k_\mathrm{B}T}}\right)
\]
is only proportional to $\lim_{\omega\to
0}\omega(1-e^{-\frac{h\omega}{2\pi k_\mathrm{B}T}})\propto \omega^2$ while the
mainly used BC spectral shape
\citep{1976CaJPh..54..593B,1985ApJ...296..644B} is constant. In the
formula above, $\omega=2\pi\nu$ is the angular frequency, $n_A$ and
$n_B$ are the densities of the species in
$\amagat^{-1}$\qq\footnote{$1 \amagat$ corresponds to a particle
number density of $2.686755\cdot 10^{19} \cm^{-3}$}, while $T$
denotes the temperature, $c$ the speed of light and $h$ Planck's
constant.

\subsection{Neglected processes}

There are more CIA data available in the literature \citep[e.g.][for
CIA of HD/\{He,Ar,H$_2$ and
H\}]{1988JChemPhys...88.4855G,2001JChemPhys...115.5427G}, but their
maximum in absorption roughly coincides with the $\HH_2/\HH_2$ and
$\HH_2/\He$ CIA in position and height. Given the small number
fraction of D and the kind of absorption (dipole radiation), for the
abundances considered here these absorption never dominates.

\subsection{Absorption coefficients}

\begin{figure*}
  \centering
  \includegraphics[width=\textwidth]{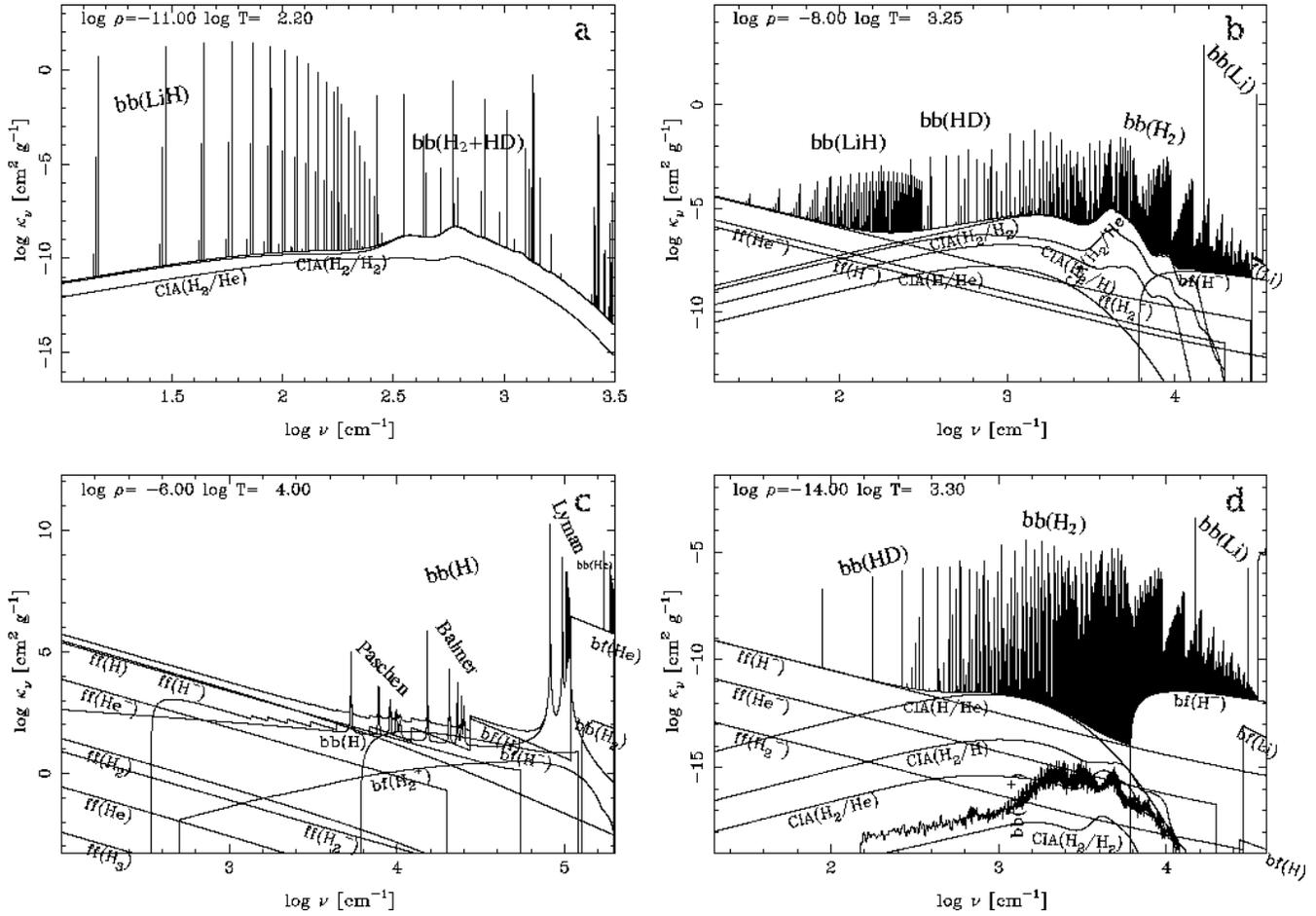}
  \caption{Absorption coefficient $\kappa_a(\nu)$ for different densities and temperatures including line absorption: {\bf (a)} $\rho=10^{-15}\g\cm^{-3}$, $T=10^{2.2}\K$; {\bf (b)} $\rho=10^{-8}\g\cm^{-3}$, $T=10^{3.25}\K$; {\bf (c)} $\rho=10^{-6}\g\cm^{-3}$, $T=10^{4}\K$; {\bf (d)} $\rho=10^{-14}\g\cm^{-3}$, $T=10^{3.3}\K$. Also shown are the contributions of the different absorption processes.}
  \label{fig:abscoef2}
\end{figure*}

\begin{figure*}

\includegraphics[width=\textwidth]{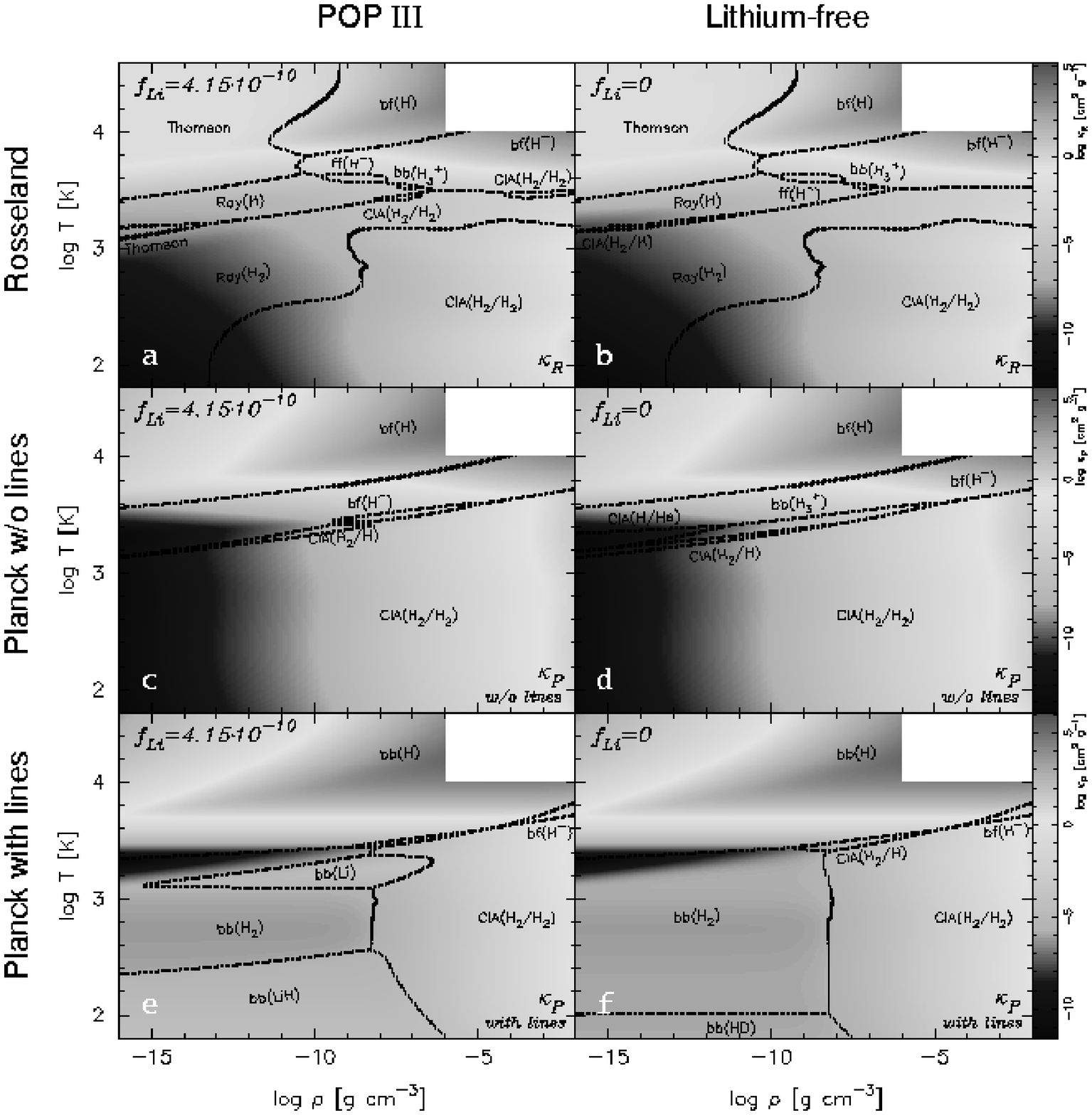}

\caption{Rosseland {\bf (a+b)} and Planck mean (without lines: {\bf
c+d}, with lines: {\bf e+f}) values for the POP {\sc III} case
($f_{\Li}=4.15\cdot 10^{-10}$, left: {\bf a,c,e}) and the zero-metal
case ($f_{\Li}=0$, right: {\bf b,d,f})} \label{fig:means}
\end{figure*}

We calculate monochromatic continuum {\bf a}bsorption and {\bf
s}cattering coefficients $\kappa_{a,i}(\nu,n_j,\rho,T)$ and
$\kappa_{s,i}(\nu,n_j,\rho,T)$ for all the 28 continuous absorption
and scattering processes (see Table \ref{tbl:absproc}) using the
cross sections $\sigma_{a/s,i}(\nu,T)$ weighted with the abundance
$n_j(\rho,T)$ of the contributing species.

For the absorption processes we correct for stimulated emission
\citep[factor $1-exp\left(-h\nu/k_\mathrm{B}T\right)$,][]{1978stat.book.....M}
unless not already included in the tabulated absorption coefficents.
We sum them both up and divide by the density $\rho$ to get the mass
absorption and scattering coefficents $\kappa_s(\nu,\rho,T)$ and
$\kappa_a(\nu,\rho,T)$ \citepalias[cf. ][]{1991ApJS...76..759L}

\begin{eqnarray}
  \kappa_{a}(\nu,\rho,T)&=&\frac{1}{\rho}\sum_i \sigma_{a,1,i}(\nu,T) n_j(\rho,T) \left(1-e^{-\frac{h\nu}{k_\mathrm{B}T}}\right) \notag \\
  &&+ \frac{1}{\rho}\sum_i \sigma_{a,2,i}(\nu,T) n_j(\rho,T)n_k(\rho,T) \left(1-e^{-\frac{h\nu}{k_\mathrm{B}T}}\right)\label{eq:nuabs}
\end{eqnarray}

\begin{eqnarray}
  \kappa_{s}(\nu,\rho,T)&=&\frac{1}{\rho}\sum_i \sigma_{s,1,i}(\nu,T) n_j(\rho,T) \notag \\
  &&+\frac{1}{\rho}\sum_i \sigma_{s,2,i}(\nu,T) n_j(\rho,T) n_k(\rho,T)\label{eq:nuscatter}
\end{eqnarray}
The indices ''1'' and ''2'' indicate 1- and 2-body absorptions.

Figs.~\ref{fig:abscoef1} and \ref{fig:abscoef2} show sample
absorption coefficients.

\section{Rosseland and Planck mean values} \label{Sect:rossplanck}

Using them together with the integrated line absorption coefficients
(\ref{eq:integabscoef}) calculations of the Rosseland and Planck
mean opacity are done according to

\begin{equation}
  \frac{1}{\kappa_R(\rho,T)} = \frac{3\pi}{4\sigma T^3} \int_0^\infty \frac{1}{\kappa_s+\kappa_a}\frac{\partial B_\nu}{\partial T} d\nu  \label{eq:ross}
\end{equation}

\begin{equation}
 \kappa_P(\rho,T) = \frac{1}{\sigma T^4} \int_0^\infty \kappa_a B_\nu d\nu +\sum_i \kappa_{P,L_i}\label{eq:planck}
\end{equation}
where the factors in front of the integrals are the normalization
conditions.

The typical range of our frequency grid is $10\dots 10^7 \cm^{-1}$
which corresponds to photon energies of $1.24\cdot(10^{-3}\dots
10^3) \eV$. We use typically $20000$ logartithmically equidistant
grid points with additional points added for a better resolution of
bound-free absorption edges.

In this contribution, calculations of Planck means are carried out
for two sets: one set only including the continuum contribution and
one including all contributions added via the integrated absorption
coefficient. The only exceptions from this scheme are $H_3^+$, CIA
and Stark broadening of atomic H. The latter obviates opacity
calculation for $\rho>10^{-5}\g \cm^{-3}$ and $T>10^{4}\K$.

The two Planck sets are two extreme cases, where the complete Planck
opacity set (continuum+lines) can be used for ``zeroth-order''
calculations, while the Planck continuum is suitable for more
detailed modeling (including lines depending on the motion of the
gas, etc.).

\subsection{Rosseland mean}

The upper part of Fig. \ref{fig:means} shows the Rosseland means for
our POP {\sc iii} Li content $f_{\textup{Li}}=4.15\cdot 10^{-10}$
and the Li-free composition.

The $\rho$-T-plane can be divided into 4 regions split by the lines
$\rho\approx 10^{-9}\g\cm^{-3}$ and $T\approx 3000\K$. At low
density scattering dominates while true absorption is dominating the
high-density region. For scattering we distinguish between Rayleigh
scattering and Thomson scattering (see Figs. \ref{fig:abscoef1}a+c)
in the low- and high-temperature regimes. True absorption is divided
into mainly bound-free and CIA absorption (see Figs.
\ref{fig:abscoef1}d+b).

In the mid-temperature region there is a steep transition between
small and large opacity values at low and high temperatures,
respectively. For the low-density region the steep gradient is
simply produced by the sharp decline/appearance of $\HH_2$, $\HH$
and $\e$ (evident in the dominance of scattering involving these
species) while at high densities bound-free absorption of $\HH^-$ is
the dominant mechanism which itself exhibits a strong temperature
dependence.

\begin{figure}
  \includegraphics[width=0.475\textwidth]{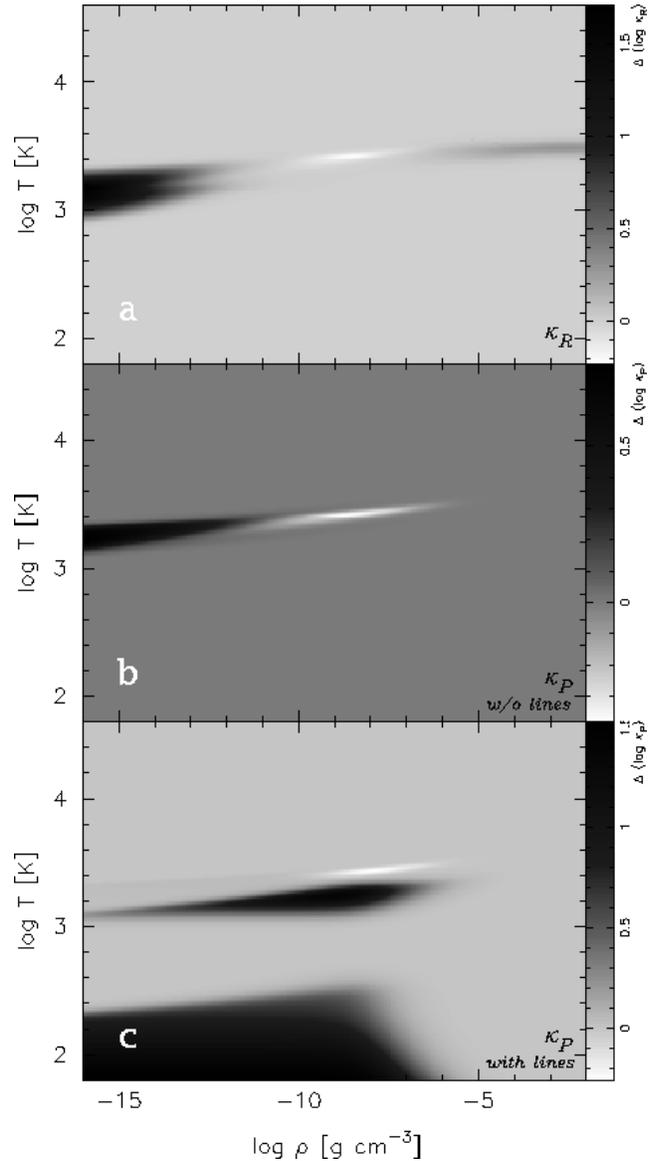}

\caption{Difference between POP {\sc III} ($f_{\Li}=4.15\cdot
10^{-10}$) and the Lithium-free case ($f_{\Li}=0$) for Rosseland and
Planck mean opacities: (a) Rosseland mean; (b) Planck mean for
continuum case; (c) Planck mean with lines} \label{fig:licomp}
\end{figure}

A comparison of the Li-free with the POP {\sc iii} case shows the
importance of taking an even minute Li contents into account (as was
already indicated by the number densities in Fig.
\ref{fig:liabundance}). The early ionisation of $\Li$ increases the
$\HH^-$ abundance which at high densities shows up as an extension
of the $\HH^-$ dominated opacity area and as an increase in the
opacity value. At lower densities the free electrons increase the
opacity due to Thomson scattering. The presence of Li, however,
decreases the Rosseland means at intermediate densities due to the
destruction of $\HH_3^+$ ions. (cf., Figs. \ref{fig:means}a and b).
In Fig. \ref{fig:licomp}a we show these differences quantitatively.

\subsection{Planck mean}

In panels c--f of Fig. \ref{fig:means} we show the Planck mean
opacities for the Pop {\sc iii} chemical composition (panels c and
e) and for the Li-free mixture (d and f). In panels c and d we show
the values for continuum absorption only, while in panels e and f
lines are taken into account.

\subsubsection{Continuum absorption}

For continuum absorption the Planck means show a rather simple
three-part structure approximately following the change from
molecular to atomic to ionic H, respectively. In the molecular H
region the only dominant true absorption is CIA of $\HH_2/\HH_2$
pairs, whereas for atomic H bound-free absorption of $\HH^-$ and for
ionised H bound-free absorption of $\HH$ are the dominant absorption
mechanisms.

Li influences the opacity in a small stripe at the boundary between
the CIA ($\HH_2/\HH_2$) and the bound-free $\HH^-$ absorption where
the influence of CIA ($\HH/\He$) and bound-bound of $\HH_3^+$
decreases. The overall difference is similar to the Rosseland means,
although the positive deviations are smaller, while the negative
ones are much larger. At this transition there is a local minimum of
the Planck mean, as $\HH_2$ dissociates/forms, while $\HH$ and
$\HH^-$ has not fully been formed/destroyed yet.

\subsubsection{Line Absorption}

Taking into account line absorption changes the result considerably:
At regions where H is mostly in its atomic or ionic form,
bound-bound transitions of atomic H yield the largest contribution
to the opacity (see Fig. \ref{fig:abscoef2}c). At lower
temperatures, the opacity contribution of CIA ($\HH_2/\HH_2$) only
remains dominant for high densities ($\rho > 10^{-8} \g\cm^{-3}$),
whereas at lower densities line absorption is the dominant mechanism
(see Figs. \ref{fig:abscoef2}a,b,d). With line absorption the local
minimum of the opacity prominently shows up: The drop in opacity
reaches 7 orders of magnitudes within a very small temperature
interval (at $T\approx 2000\K$) and low densities.

D is an important absorber only for temperatures below $\approx
100\K$, as then the lowest-lying rotational transition of $\HD$ at
$\approx 128\K$ outperforms the $\HH_2$ transition at $510\K$ in its
contribution to the Planck mean.

The direct influence of Li on the Planck means comes into play
through the absorption of atomic Li (cf. Fig. \ref{fig:abscoef2}b)
and $\Li\HH$ (Li hydride, see Fig. \ref{fig:abscoef2}a). Although
the number density of these species is much smaller than those of D
(itself being of considerably smaller abundance than H and He), they
increase the opacity by one and a half orders of magnitude. The
indirect influence of the $\HH^-$ absorption and $\HH_3^+$
destruction is still present.

Atomic Li absorbs mainly through the 6708 $\AA$ transition. It is
stronger than the quadrupole absorption of $\HH_2$ through its large
Einstein coefficient despite taking place at a comparatively long
wavelength. However the influence of this absorption ceases with
increasing temperature where Li gets ionized.

For $\LiH$, the Einstein coefficient is larger than that of $\HH_2$
transitions due to the presence of a strong dipole moment, but it
does not reach the values of the 6708 $\AA$ transition of atomic Li.
$\LiH$ becomes dominant due to the smaller energy differences
spacings between its rotational levels. This considerably reduces
the wavelength of the base transition. Thus the Planck function for
temperatures lower than $510 \K$ (lowest rotational state of
$\HH_2$) still catches rotational and vibrational transitions of
$\LiH$ (lowest rotational state at $21 \K$) while the $\HH_2$
contribution decreases.

The double-peaked CIA dominated opacities at temperatures below
$1000 \K$ can be attributed to the the isotropic overlap induced
dipole and the quadrupolar field induced dipole via isotropic
polarizability components in the fundamental band of $\HH_2/\HH_2$
\citep[][cf., our Fig.
\ref{fig:abscoef1}]{1996Icar..123....4B,1989PhRvA..39.2434M}.

\subsection{The quantitative influence of Lithium on the
opacity} \label{Sect:quantLithium}

Here, we summarize the differences the presence or absence of Li in
the chemical composition of the matter makes. In the Rosseland and
Planck means Li manifests itself by
\begin{itemize}
\item its early ionisation;
\item the destruction of $\HH_3^+$ in its presence;
\item the $6708\AA$ bound-bound transition of atomic Li;
and by
\item the molecular absorption of $\LiH$.
\end{itemize}

\noindent The first two mechanisms show up in both the Rosseland and
Planck means, while the latter only appears in the Planck line
opacity. All mechanisms except the $\HH_3^+$ destruction lead to an
increase of the opacity.

\begin{figure}
  \centering
  \includegraphics[height=0.475\textwidth,angle=-90]{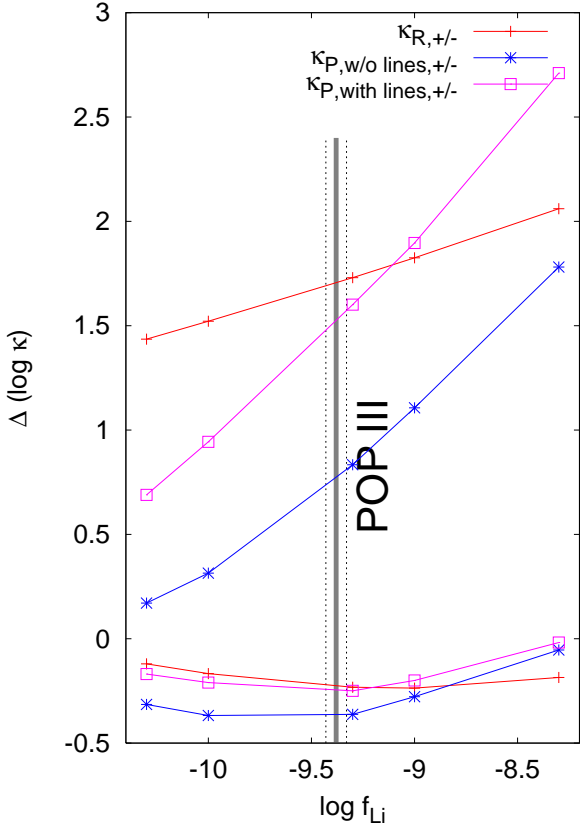}
  \caption{Deviations from Li-free ($f_{\Li}=0$) opacity for
  different Li contents. $\Delta\left(\log\kappa\right)>0$ indicates
  positive deviations, while $\Delta\left(\log\kappa\right)<0$
  indicates negative deviations relative to Li-free opacity. The
  fiducial POP {\sc{iii}} content is given including the uncertainties.}
  \label{fig:li-quant}
\end{figure}

In Fig. \ref{fig:li-quant} we show for different Li fractions ($\log
f_\mathrm{Li} = -10.3 \dots -8.3$) the differences from Li-free
composition.

The positive deviations ($\Delta \log \kappa>0$) show a steady
increase with increasing Li content. The negative deviations
(predominantly due to destruction of $\HH_3^+$) increase for Li
fractions $f_{Li}<5\cdot 10^{-10}$, but are being drowned by the
increase of opacity for larger Li fractions.

\section{A more detailed assessment of the chemical equilibrium}
\label{Sect:chemequilibrium}

As we have seen, at low temperatures $\HH_2$ is the main absorber.
Due to the lack of dust which would act as a catalyst for $\HH_2$
formation \citep[for a comparison, see][]{2003ApJ...584..331G},
$\HH_2$ formation and therefore cooling and absorption is small. The
main $\HH_2$ formation channels are via $\HH^-$ and $\HH_2^+$ at low
and high temperatures, respectively \citep[e.g.
][]{1998A&A...335..403G,2002JPhB...35R..57S}. These 2-body reactions
operate on rather long timescales. However at higher densities
($n\ge 10^8 \cm^{-3}$) 3-body reactions become efficient
\citep{1983ApJ...271..632P}.

In the following, we look into the timescales for molecule formation
in order to assess where these chemical timescales exceed the
hydrodynamical one (for which we take the free-fall time scale).
Thus we can estimate at which densities and temperatures the
assumption of chemical equilibrium is satisfied at all.

For all timescale calculations we assume that initially the
constituents of the molecules are mainly in their neutral atomic
form and subsequently get converted. This approach can be justified
given the fact that in the cosmological fluid H and He are in their
neutral atomic form starting from a redshift of $z\approx 1100$
while for Li we take $z\approx 100$
\citep[][]{1998A&A...335..403G,2002JPhB...35R..57S}.

\subsection{Molecular Hydrogen $\HH_2$}

The $\HH_2$ formation timescale $\tau_{\HH_2}$ due to the reaction
$\HH+\HH+\HH \rightarrow \HH_2+\HH$ can be estimated using $k_4$ of
\citep{1983ApJ...271..632P} to
\begin{equation}
  \label{eq:tauh2}
  \tau_{\HH_2}=2420 \yr \cdot \left(\frac{T}{300\K}\right)\left(\frac{\rho}{10^{-14}\g\cm^{-3}}\right)^{-2}
\end{equation}

In contrast to the rapid 3-body-reaction the only relevant ionic
$\HH_2$ formation channel at low temperatures is via $\HH^-$ while
at higher temperatures $\HH_2$ is formed via $\HH_2^+$
\citep[][]{1961Obser..81..240M,1968ApJ...154..891P,1967Natur..216..967S}.
The $\HH^-$-consuming reaction is $\HH^- + \HH\rightarrow \HH_2+\e$.
This reaction depends on the $\HH^-$ abundance which never is the
most abundant H species and rapidly reaches chemical equilibrium as
$\HH^-$ abundance determining reactions proceed on timescales much
shorter than those for the $\HH_2$ chemistry
\citep[][]{1997NewA....2..181A}. Thus taking into account the
$\HH_2$ formation rate for the ionic reaction channel is not
advisable, as it would only give the timescale on which all $\HH^-$
is depleted. The maximum $\HH_2$ abundance due to this reaction
never exceeds the available $\HH^-$ density.

\begin{figure}
  \centering
  \includegraphics[height=0.475\textwidth,angle=-90]{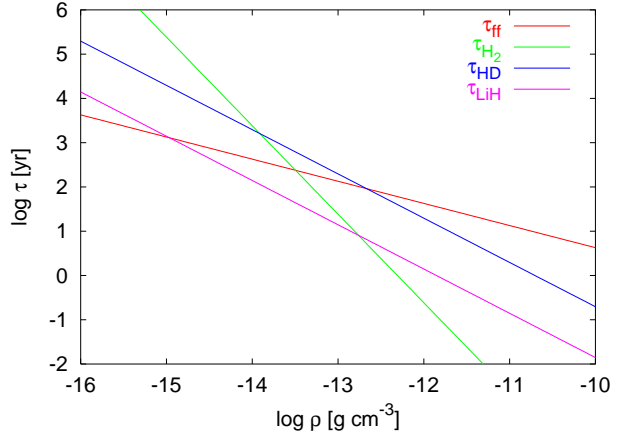}
  \caption{Comparison of free-fall timescale with the molecule formation timescales at $T=300\K$ and a $\HH_2$ fraction of $10^{-5}$}
  \label{fig:timescale}
\end{figure}

Following arguments similarly shown by \cite{2004NewA....9..353B},
there exists a critical density (as a function of temperature), at
which $\HH_2$ formation proceeds within one free-fall time
\[
t_{\ff}=\sqrt{\frac{3}{8\pi G \rho}}=425\yr\cdot \left(\frac{\rho}{10^{-14}\g\cm^{-3}}\right)^{-\frac{1}{2}}
\]

\[
\rho_{\textup{crit,$\HH_2$}}=3.2\cdot 10^{-14} \g\cm^{-3}\cdot \left(\frac{T}{300\K}\right)^\frac{2}{3}\g \cm^{-3}
\]

\subsection{Deuterium hydride $\HD$}

$\HD$ is formed mainly via deuteron exchange with $\HH_2$
($\HH_2+\DD\rightarrow\HD+\HH$) For the timescale we use rate
coefficient (1) of \citet{2002P&SS...50.1197G} and get for the
timescale (setting $T=300\K$)

\begin{equation}
\tau_{\HD}=1.97 \cdot 10^{3}\yr\cdot
\left(\frac{\rho}{10^{-14}\g\cm^{-3}}\right)^{-1}\left(\frac{f_{\HH_2}}{10^{-5}}\right)^{-1}\nonumber
\end{equation}
Note that $\tau_{\HD}$ depends on the $\HH_2$ fraction. Thus with
increasing $\HH_2$ abundance this timescale is reduced.

\subsection{Lithium hydride $\LiH$}

For LiH we use the low-temperature rate of reaction (20) of
\citet{1996ApJ...458..401S} and get
\begin{equation}
   \tau_{\LiH}=140\yr\cdot \left(\frac{\rho}{10^{-14}
   \g\cm^{-3}}\right)^{-1}\left(\frac{T}{300\K}\right)^{-0.28}\nonumber
\end{equation}
and
\[
\rho_{\textup{crit, $\LiH$}}=1.08\cdot 10^{-15} \left(\frac{T}{300\K}\right)^{-0.56}\g \cm^{-3}
\]
The situation, however, will change, once there are data available
of $\LiH$ producing 3-body reactions \citep[see discussion
in][]{1996ApJ...458..401S}. This will modify the density dependence
of the formation timescale ($\tau\propto \rho^{-2}$). The same
applies for $\HD$.

At densities $\rho > 10^{-14} \g\cm^{-3}$ chemical equilibrium of
$\HH_2$ is reached within one free-fall time scale or less. Thus,
for our the computational domain
($10^{-16}\g\cm^{-3}<\rho<10^{-2}\g\cm^{-3}$) the assumption of
chemical equilibrium is justified, with the exception of the two
lowest-density decades (which we tabulated nevertheless for
computational convenience). $\LiH$ can be formed even more rapidly
and thus can be regarded as being in equilibrium for densities $\rho
> 10^{-15}\g\cm^{-3}$. $\HD$ is rather hard to form, as long as
there is no significant $\HH_2$ abundance because the formation
timescale crucially depends on the $\HH_2$ fraction. However, we
have seen in Sect. \ref{Sect:rossplanck} $\LiH$ to be a significant
absorber while $\HD$ is not. Thus the $\HD$ abundance is of only
minor relevance for the opacities presented.

$\tau_{\HD}$, together with the other timescales discussed in this
Section, is shown in Fig. \ref{fig:timescale}.

\section{Comparison with existing calculations}\label{Sect:compare}

In order verify our results, we compare the Rosseland means of our
calculation with existing calculations of $Z=0$ opacities by setting
$f_{\Li}=f_{\DD}=0$. We restrict the comparison to more recent
opacities (Section \ref{Sect:sps86}-\ref{Sect:Harris04}). A common
difference of all these opacities compared to our data can be found
for temperatures around $10^{3.2\dots 3.4}\K$. This comes about due
to our use of the newly calculated $\HH_3^+$ equilibrium constant in
conjunction with the symmetry factor in the $\HH_2$ dissociation
equilibrium.

An overview of previous low metallicity opacities is shown in Table
\ref{tbl:overview}.

\begin{table*}
   \centering
 \renewcommand{\arraystretch}{2.0}
   \begin{tabular}{l|c|c|c|l}
     \hline
     \hline
     Reference & Opacity type & Density range & Temperature range & Chemical
     composition\\
     \hline
     \hline
     \citet{1976ApJS...31..271C} Paczynski I-IV mixture & $\kappa_R$ & $-12<\log \rho<-4$ & $1500<T<12000\K$ & $X=0.0\dots 1.0$\\
     \hline
     \citet{1986ApJ...302..590S} & $\kappa_R$ & $-15<\log \rho < -3$ & $1500<T<7000\K$ & $X=0.72$\\
     \hline
     \citet{1991ApJS...76..759L} & $\kappa_{R/P}$ & $-12<\log \rho < -0.3$ & $1000<T<7000\K$ & $X=0.72$\\
     \hline
     \citet{1994ApJ...437..879A} & $\kappa_{R/P}$ & $-7<\log R<1$ & $3.0<\log T<4.1$ & $X=0.0\dots 0.8$\\
     \hline
     \citet{1994MNRAS.266..805S} OP Project & $\kappa_{R/P}$ & $-7<\log R<-1$ &$3.5<\log T<8.0$& $X=0.0\dots0.7875$\\
     \hline
     \citet{1996ApJ...464..943I} OPAL & $\kappa_R$ & $-8<\log R<1$ & $3.75<\log T<8.70$ & $X=0.0\dots 1.0$\\
     \hline
     \citet{2004ApJ...600.1025H} & $\kappa_R$ & $-14<\log \rho < -2$ & $1000<T<9000\K$ & $X=0.0\dots 1.0$\\
     \hline
     This work & $\kappa_{R/P}$ & $-16<\log\rho<-2$ & $1.8<\log T<4.6$ & $X=0.75$\\
     \hline
     \hline
   \end{tabular}
   \caption{Characteristics of $Z=0$ opacity calculations or opacity calculations with a $Z=0$ subset ($\rho$ in g\,cm$^{-3}$; $R=\rho T_6^{-3}$; $T$ in K; $T_6 = T /
   10^6\,\mathrm{K}$)}
   \label{tbl:overview}
 \end{table*}

\subsection{\citet{1986ApJ...302..590S}} \label{Sect:sps86}
\begin{figure*}
\centering\includegraphics[width=0.725\textwidth]{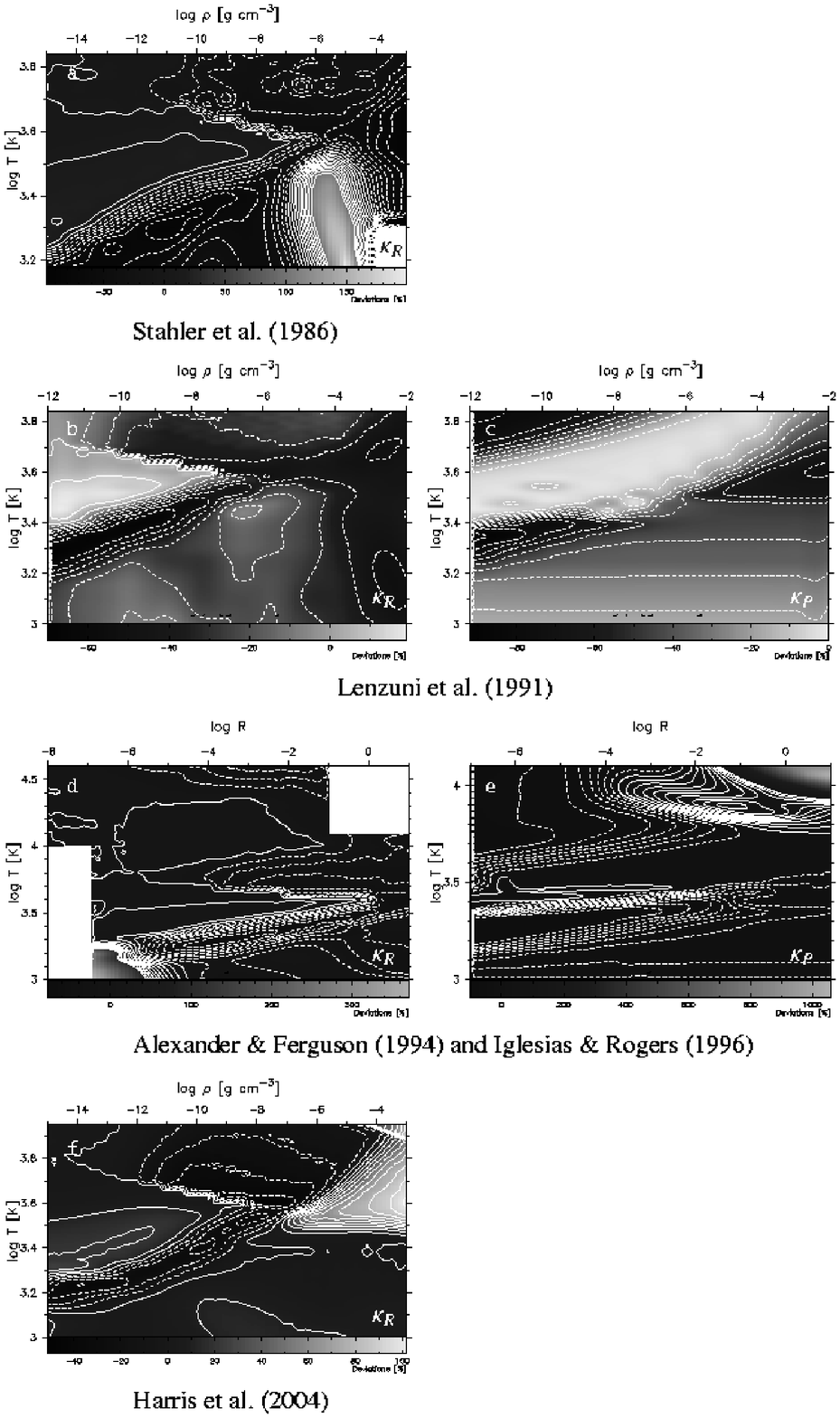}
\caption{Comparison of our Rosseland and Planck mean opacities with
previously published $Z=0$-opacities: {\bf (a)} $\kappa_\textrm{R}$
by \citet{1986ApJ...302..590S} ($X=0.72$, $Y=0.28$, and $Z=0$); {\bf
(b) \& (c)} $\kappa_\textrm{R}$ and $\kappa_\textrm{R}$ by
\citetalias{1991ApJS...76..759L} ($X=0.70$, $Y=0.30$); {\bf (d) \&
(e)}  $\kappa_\textrm{R}$ and $\kappa_\textrm{P}$ by
\citet{1994ApJ...437..879A} for $\log T<4$ ($\log T < 4.1$ for
Planck mean) and \citet{1996ApJ...464..943I} for $\log T>4$
($X=0.70$, $Y=0.30$); {\bf (f)} $\kappa_\textrm{R}$ by
\citet{2004ApJ...600.1025H} ($X=0.72$, $Y=0.28$). The contour lines
correspond to differences of the opacity in \%. Dashed lines
indicate negative, solid lines positive differences relative to our
opacities. The increment between neighbouring contours is 10 \%.}
\label{fig:comparisons}
\end{figure*}

The match with the \citetalias{1986ApJ...302..590S} opacities
(Fig.~\ref{fig:comparisons}a) is good as long as $H^-$ is the
dominant absorption mechanism. For low temperatures, the deviations
are due to the use of CIA data of \citet{1969ApJ...156..989L} and
\citet{1971JQSRT...11..1331P} while we use the more recent Borysow
data. The deviations at higher temperatures and densities are due to
the influence of the Stark broadening of H lines which have been
neglected in the \citetalias{1986ApJ...302..590S} calculation.

\subsection{\citet{1991ApJS...76..759L}} \label{Sect:lenzuni91}

The comparison with \citet{1991ApJS...76..759L} is done for both
Rosseland and Planck means (Fig.~\ref{fig:comparisons}b\&c).

At higher temperatures and densities the deviations in the Rosseland
mean are again due to Stark broadened H lines. For lower
temperatures the deviations are only in the 20-30 \% range, much
smaller than with respect to \citetalias{1986ApJ...302..590S}. This
is due to their use of the partially newly available CIA data of
$\HH_2/\HH_2$ and $\HH_2/\He$ collisions for the roto-vibrational
and roto-translational transitions while still having to use
extrapolations for the overtones \citet[using data
by][]{1969ApJ...156..989L}.

For the Planck means the differences are considerably higher. This
is due to the linearity of Planck averaging which prefers peaks in
the opacity (CIA at low temperatures, H lines at higher temperatures
and lower densities).

\subsection{\citet{1994ApJ...437..879A} \& \citet{1996ApJ...464..943I}
Z=0 sets} \label{Sect:G93}

The comparison to the OPAL opacities for the Grevesse \& Noel
chemical composition at Z=0 shows a much better concordance over a
larger temperature range (Fig. \ref{fig:comparisons}d\&e).

In the high-temperature limit ($\log T > 3.8$) the Rosseland mean
opacities are consistent at the 10\% level (Stark broadening). For
the very highest temperatures, deviations are due to the
approximation of the statistical weight of H by the ground state
which no longer is true. At lower temperatures we again have a
10-20\% deviation due to CIA.

While the inclusion of Stark broadening for the Rosseland means
shows a good consistence with the OPAL data, there are deviations in
the Planck means. These differences are due to our coarse frequency
grid which does not correctly trace all the peaks of the Hydrogen
lines.

\subsection{\citet{2004ApJ...600.1025H}} \label{Sect:Harris04}

The comparison to the \citetalias{2004ApJ...600.1025H} data in Fig.
\ref{fig:comparisons}f shows an overall good agreement with our
data. The discussion is analogous to the OPAL comparison. except
that now we are consistent at the 10\% level accuracy for lower
temperatures due to their use of $\HH/\He$, $\HH_2/\HH$ and more
recent $\HH_2/\HH_2$ and $\HH_2/\He$ CIA data.

At densities $\rho > 10^{-7}\g\cm^{-3}$ and temperatures $\log T
> 3.5$ there are, however, important differences. These are due to
\citetalias{2004ApJ...600.1025H}'s use of two different approaches
to calculate the $H_3^+$ abundance in order to overcome machine
accuracy problems. In their $\HH_2$ dominated regime, they take a
dissociation energy of $4.52 \eV$ whereas we take $4.478 \eV$
\citep[following][]{huberherzberg}. In our calculations, we do not
find any problems regarding machine accuracy. The second reason for
the deviations is probably due to the difference in $\HH_2^+$
absorption coefficients where we use the \citet{1994ApJ...430..360S}
data while they use \citet{2000ARep...44..338L}.

\section{Conclusions} \label{Sect:conclusion}

Rosseland and Planck mean opacities for primordial matter have been
calculated using the most recent available data for the absorption
mechanisms.

We tabulate the opacity data in Tables~E1, E2 and E3, for a temperature range
$1.8<\log ( T / \mathrm{K}) <4.6$ and $-16 < \log ( \rho /
\mathrm{g\,\cm}^{-3}) < -2$ for our POP {\sc iii} matter composition
(cf. Table \ref{tbl:abundances}). These are the first POP {\sc iii}
opacity tables for temperatures $T < 1000\K$.

In order to make application as easy as possible, we provide two
sets of Planck opacities: Planck means including only continuum
absorption and those including molecular lines. Higher resolution
tables including routines for bicubic interpolation tables are
available from the authors upon request.

It has been shown (see Sect. \ref{Sect:rossplanck} and
\ref{Sect:quantLithium}) that the small number fraction of Li leads
to a significant change in the opacity values at temperatures $T <
4000\,$K compared to a pure H/He mixture. The differences can reach
up to 2 orders of magnitudes. There are four processes which change
the opacity:

As an alkali metal, Li gets ionized at comparatively low
temperatures. It therefore increases the number of available
electrons considerably. They increase the Thomson scattering
contribution to the Rosseland mean at low densities while increasing
the $\HH^-$ absorption at higher densities. For both the Planck
continuum and line case only the $\HH^-$ absorption at low densities
(no scattering) is enhanced.

In the presence of metals (e.g. Lithium) $\HH_3^+$ is being
destroyed. As $\HH_3^+$ is the most abundant positive ion at
temperatures around $3000\K$ and densities $\rho >
10^{-12}\g\cm^{-3}$ it influences the opacity indirectly in changing
the chemical equilibrium when being destroyed. Furthermore, in part
it shows up directly via bound-bound transitions of $\HH_3^+$ at
intermediate densities. For higher densities this is being drowned
by the indirect increase of the $\HH^-$-absorption.

The Planck means including line absorption are changed due to
absorption by atomic Li and molecular LiH.

For atomic Li the most important transition is the $6708\AA$
feature. The Planck function, however, is most sensible to features
at $6708\AA$ for temperatures of approx. $7000 \K$, a temperature at
which Li is ionized, at the latest. Absorption via this transition
is still important at temperatures around $1500\K$ because the
corresponding Einstein coefficient is very large compared to the
quadrupole transitions of molecular and atomic H.

At temperatures $T < 500\K$ the influence of $\HH_2$ is ceasing as
the lowest lying transition of $\HH_2$ has an equivalent temperature
of $512\K$. For lower temperatures either $\HD$ or $\LiH$ contribute
to the opacities. \LiH, however, has got a larger dipole moment and
thus has got much larger Einstein coefficients and transitions at
much lower temperatures (as low as $15\K$).

A critical point in primordial chemistry is the formation of
molecules. Formation times have been calculated for $\HH_2$, $\HD$
and $\LiH$. For densities $\rho \gtrsim 10^{-14}\g\cm^{-3}$ molecule
formation proceeds within one free-fall time (except $\HD$, which
does not play any significant role in the POP {\sc iii} case). Hence
the calculation is valid for densities larger than that, provided
the free-fall time scale being the shortest timescale relevant,
other than the chemical. We give values for densities as low as
$10^{-16}\g\cm^{-3}$ for numerical convenience.

In comparison to previous calculations we find a good agreement of
our results when neglecting the newly added absorption mechanisms
and the contributions of Li.

Based on our new opacities the influence of Li on the different
stages of POP {\sc iii} star formation and evolution can now be
assessed.

\section*{acknowledgements}
The authors acknowledge support from the {\it Deut\-sche
For\-schungs\-ge\-mein\-schaft, DFG\/} through grant {\it SFB 439
(A7)\/}. Part of this work was carried out while one of the authors
(MM) stayed as an EARA Marie Curie Fellow at the Institute of
Astronomy, University of Cambridge, UK. The hospitality of the IoA,
and of Prof. J.E. Pringle in particular, is gratefully acknowledged.
The authors thank Profs. Wehrse and Gail for valuable discussions on
the subject of this paper, We are grateful to Dr. Evelyne Roueff for
providing the HD transition probabilities. This work has made
extensive use of NASA's Astrophysics Data System.


\appendix

\section{EOS for Pop {\sc iii}} \label{App:EOS}

Primordial matter is assumed to consist of a number fraction
$f_\mathrm{E}$ of element E. Depending on the density, we define the
number densities of the involved elements
\[
N_{\{\HH, \He, \DD, \Li\}} =
\frac{\{f_\HH,1-f_\HH-f_\DD-f_{\Li},f_\DD,f_{\Li}\}}{4-3f_\HH-2f_\DD+f_{\Li}}\frac{\rho}{m_p}
\]

\noindent We need 4 equations (\ref{eq:h}--\ref{eq:li}) to describe
the conservation of the number densities of the elements (mass
conservation) and an additional one (\ref{eq:charge}) to take care
of charge neutrality. The number density of species X is called
$\left[\mathrm{X}\right]$.
\begin{eqnarray}
\conc{\HH^-}+\conc{\HH} + \conc{\HH^+} +3\conc{\HH_3^+}+ 2\left( \conc{\HH_2} + \conc{\HH_2^+}\right) &=&N_\HH \label{eq:h}\\
\conc{\He}+\conc{\He^+} + \conc{\He^{++}}+2\conc{\He_2^+} &=& N_{\He}\\
\conc{\HD}+\conc{\DD}+\conc{\DD^+} +\conc{\HH_2\DD^+}&=& N_{\DD} \\
\conc{\LiH}+\conc{\Li}+\conc{\Li^+}+\conc{\Li^{++}} &=& N_{\Li} \label{eq:li}\\
\conc{\e}+\conc{\HH^-}-\conc{\HH^+}-\conc{\HH_2^+}-\conc{\HH_3^+}-\conc{\He^+}&& \notag\\
-2\conc{\He_2^+} - \conc{\DD^+} - \conc{\HH_2\DD^+} - \conc{\Li^+} -
\conc{\Li^{++}} &=& \label{eq:charge}0
\end{eqnarray}
Note that we neglect the H contribution of the D and L species in
the H sum. Given the relative number fraction of D and Li with
respect to H this assumption is satisfied. The equilibrium constants
are given in Table \ref{tbl:eqconst}.

For the A $\lr$ B+C Saha equilibria we use equations of the form
\[
 ^nK(T)= \frac{Q_\mathrm{B} Q_\mathrm{C}}{Q_\mathrm{A}}  \left(\xi T\right)^\frac{3}{2}e^{-\frac{U}{kT}}
\]
with $\xi_{\textup{ion}} = 2\pi m_\e k_\mathrm{B} / h^2 = 1.80\cdot
10^{10}\K^{-1} \cm^{-2}$ and $\xi_{\textup{diss}}= 2\pi m_\mathrm{p}
k_\mathrm{B} / h^2 =3.30\cdot 10^{13}\K^{-1} \cm^{-2}$. $U$ is the
characteristic energy for ionisation or dissociation, $Q$ the
partition function of the contributing species. For
$\HH_2$-dissociation we use $\xi_{\textup{diss}}$ and correct the
equilibrium constant with the symmetry factor for a homonuclear
molecule \citep{1966PDAO...13....1T}.

For the solution of the system of equations we define a critical
Temperature $T_\mathrm{crit}=200 \K$ above which we solve the full
system including the charge neutrality condition. We call this the
\textit{ionic limit\/}. Below $T_\mathrm{crit}$ we neglect all ions
and solve (\ref{eq:h}-\ref{eq:li}) in the molecular limit.

In the ionic limit, we start with an initial guess of $\conc{\e}$,
solve eqns. (\ref{eq:h}-\ref{eq:li}) to derive a new $\conc{\e}$
from the neutrality condition eq. (\ref{eq:charge}). Iteration is
done with $\conc{\e}\rightarrow \sqrt{\conc{\e}_\mathrm{new}
\conc{\e}_\mathrm{old}}$. After 20-50 iterations convergence is
reached. The solution of eqns. (\ref{eq:h}-\ref{eq:li}) is done
analytically and would involve a cubic equation in the case of eq.
(\ref{eq:h}), otherwise only quadratic equations are involved. The
cubic term in the H species equation arises due the inclusion of the
$H_3^+$ abundance. As this species never is the main contributor to
the number density, we neglect it in the number density
conservation. Note that we also neglect the H containing D and Li
species in the H conservation equation. Since we have $f_\DD / f_\HH
\ll 1$ and $f_{\Li}/f_\HH \ll 1$ this assumption is justified.

In the molecular limit we have to solve analytically only eqns.
(\ref{eq:h}-\ref{eq:li}) neglecting ions.

\section{The integrated absorption coefficient} \label{Sect:integabs}

The absorption coefficient $\alpha_\nu$ $\left[\cm^{-1}\right]$ for
the $2\rightarrow 1$ transition can be written as
\[
\alpha_\nu=\frac{c^2}{8\pi\nu^2}\Phi(\nu) n_1
A_{12}\left(1-e^{-\frac{h\nu_{12}}{k_\mathrm{B}T}}\right),
\]
where $\Phi(\nu)$ the profile function, $n_1$ is the number density
of atoms in state 1, $A_{12}$ the Einstein coefficient and
$\nu_{12}$ the transition frequency. In relation to the ground state
we have
\[
\alpha_\nu=\frac{g_1}{g_0}\frac{c^2}{8\pi\nu^2}\Phi(\nu) n_0
A_{12}\left(e^{-\frac{E_1}{k_\mathrm{B}T}}-e^{-\frac{E_2}{k_\mathrm{B}T}}\right),
\]
where $E_1$ and $E_2$ are the energies of the states relative to the
ground state. In terms of the total number density $n=\sum_{i} n_i =
\left( n_0 / g_0 \right) \sum_{i} g_i \exp(- E_i / k_\mathrm{B}T ) = n_0 Q(T) /
g_0$
\[
\alpha_\nu=g_1\frac{c^2}{8\pi\nu^2 Q(T)}\Phi(\nu) n
A_{12}\left(e^{-\frac{E_1}{k_\mathrm{B}T}}-e^{-\frac{E_2}{k_\mathrm{B}T}}\right).
\]
Via $g_1 A_{12}=g_2 A_{21}$ we easily can transform to the usually
used Einstein coefficient for spontaneous emission, $A_{21}$.

Evaluating $\int_0^\infty \alpha\nu d\nu$ yields
($\int_0^\infty\Phi(\nu) d\nu=1$) the integrated absorption
coefficient $\chi_{\nu_{12}}$ $\left[\cm^{-1}\s^{-1}\right]$
\[
\chi_{\nu_{12}}=g_1\frac{c^2}{8\pi\nu_ {12}^2 Q(T)} n
A_{12}\left(e^{-\frac{E_1}{k_\mathrm{B}T}}-e^{-\frac{E_2}{k_\mathrm{B}T}}\right),
\]
where we made use of the approximation $\nu\approx \const$ across
the line.

For Planck averaging we furthermore assume $B_\nu\approx \const$
across the line. Then the line contribution of the $2\rightarrow 1$
transition is
\begin{eqnarray}
\kappa_{P,L}(\rho,T)&=&\frac{1}{\rho}\frac{\int_\nu \alpha_\nu B_\nu
d\nu}{\int_\nu B_\nu d\nu}\notag\\
&=& \frac{1}{\sigma \rho T^4} \int_\nu \alpha_\nu B_\nu d\nu\notag
\\
&=& \frac{n}{\sigma \rho T^4} \sum_{\nu_{1\rightarrow 2}}
\chi_{\nu_{12}} B_{\nu_{12}} \notag\\
&=& \frac{n}{\sigma \rho T^4} \sum_{\nu_{1\rightarrow 2}}
g_1\frac{1}{8\pi k_{1\rightarrow 2}^2 Q(T)}
A_{12}\left(e^{-\frac{E_1}{k_\mathrm{B}T}}-e^{-\frac{E_2}{k_\mathrm{B}T}}\right) B_\nu
\notag\\
&=& \frac{hcn}{4\pi \sigma \rho T^4 Q(T)} \sum_{k_{1\rightarrow
2}} g_1 A_{12}\left(e^{-\frac{E_1}{k_\mathrm{B}T}}-e^{-\frac{E_2}{k_\mathrm{B}T}}\right)
\frac{k_{1\rightarrow
2}}{e^\frac{hc\nu_{1\rightarrow 2}}{k_\mathrm{B}T}-1} \notag\\
\label{eq:integabscoef}
\end{eqnarray}
where we introduced the wavenumber $k_{1\rightarrow 2}$ corresponding to the
$1\rightarrow 2$ transition. Multiplied with $\rho / n$,
the last expression only depends on the temperature. Calculated once
for each species summing up all the lines for the temperature grid
chosen, it can be added directly to the Planck mean with the density
weight $n / \rho$.

Note that we make use of the Einstein coefficient $A_{12}$, which is
related to the Einstein coefficient for spontaneous emission
$A_{21}$ through the Einstein relation $g_1A_{12}=g_2A_{21}$.

\section{Equilibrium $\HH_2^++\HH_2\leftrightarrow \HH_3^++\HH$ }
\label{Sect:h3plus}

We have fitted the equilibrium constant $\log_{10} K$ to the 6
parameter function ($x=\log_{10} (T/ \K) $)
\begin{equation*}
  \log_{10} K(T) = \frac{7079\K}{T}\left(1+\sum_{i=1}^5 a_i x^i\right)
\end{equation*}

with the coefficients
\begin{eqnarray*}
a_1 &=& +\left(0.365793      \pm 0.02124\right)\\
a_2 &=&  -\left(0.624287        \pm 0.03488 \right) \\
a_3 &=&  +\left(0.388540          \pm 0.02095\right)\\
a_4 &=& -\left(0.102057        \pm 0.00546\right) \\
 a_5 &=& +\left(0.940462       \pm 0.05203\right)\cdot 10^{-2}
\end{eqnarray*}
This fit causes maximum deviations of 4 $\%$ of $K$ for
$300\K<T<7000\K$, whereas the maximum deviation never exceeds 8 $\%$
below $10^{4} \K$. $7079\K$ correspond to the enthalpy of $16300 \K$
for the reaction $\HH_2^++\HH_2\rightarrow \HH_3^++\HH$, divided by
the $\log 10$ factor due to basis conversion.

\section{A remark on the standard format of opacity tables} \label{Sect:tblfmt}

Opacity mean values are usually tabulated in the $(R, T)$-plane,
where $R=\frac{\rho}{T_6^3}$. $T_6$ is the temperature in units of
$10^6$ K, $\rho$ the density. Using this $R$-parameter, the tables
remain rectangular over large temperature ranges.

We have the gas ($P_\mathrm{g}$) and radiation ($P_\mathrm{r}$)
pressure (here in the optically thick case)
\begin{equation}
  \label{eq:gas+rad}
  P_\mathrm{g}=\rho \frac{kT}{\mu m_\mathrm{p}} \qquad P_\mathrm{r} = \frac{4\sigma}{3c}T^4
\end{equation}

Calculating the ratio and taking the logarithm we have
\begin{equation}
  \label{eq:xi}
  \bar \xi = \log \left(\frac{P_\mathrm{g}}{P_\mathrm{r}}\right) = \log \left(\frac{3k_\mathrm{B} c}{4\sigma\mu m_\mathrm{p}}\right) + 3 \bar T - \bar \rho \approx 22.5  + 3 \bar T - \bar \rho
\end{equation}

Subtracting the logarithmized definition of $\bar R= \bar \rho +18 - 3\bar T$, we have
\[
\bar \xi - \bar R = 4.5
\]

For $\bar R= -4.5$ gas and radiation pressure are equal. As the
range of $\bar R$ usually tabulated is $-8<\bar R<1$ (OPAL, OP
Project), $\bar R=-4.5$ is in the middle of this coordinate range.

The physical meaning of introducing $\bar R$ is that one remains in
the physical regime where gas and radiation pressure compete with
each other. Too large departures from this equality require changes
in the EOS.

Therefore the each column of opacity tables in the style of OPAL \&
OP represent different gas/radiation pressure ratios.

 \section{Sample tables}

Sample tables for Rosseland and Planck means are given in
Tables~E1, E2 and E3 (will be available in the accepted version).

\label{lastpage}

\end{document}